\documentclass{emulateapj}
\usepackage{url}

\newcommand{\photonfluxunit}{{\rm photons~cm}^{-2} {\rm~s}^{-1} {\rm~\AA}^{-1}}

\newcommand{\csfuvflux}{(4.9 \pm1.8) \times10^{-6}}% ~\photonfluxunit}
\newcommand{\csnuvflux}{(10.5 \pm 0.4) \times10^{-5}}% ~\photonfluxunit}
\newcommand{\irfuvflux}{(3.6\pm1.1) \times 10^{-5}}% ~\photonfluxunit} 
\newcommand{\orfuvflux}{(2.3\pm0.3) \times 10^{-4}}% ~\photonfluxunit} 
\newcommand{\ornuvflux}{(3.9\pm1.3) \times 10^{-5}}% ~\photonfluxunit} 
\slugcomment{Accepted by the Astrophysical Journal Letters}

\shorttitle{Extended UV Emission from U Hya}
\shortauthors{Sanchez et al.}

\begin{document}
\title{First Detection of UV emission from a Detached Dust Shell:\\GALEX Observations of the Carbon AGB Star U Hya}

\author{Enmanuel Sanchez\altaffilmark{1}}
\affil{Florida State University, Tallahassee, FL USA}
\author{Rodolfo Montez Jr.} 
\affil{Department of Physics and Astronomy, Vanderbilt University, Nashville, TN USA} 
\author{Sofia Ramstedt} 
\affil{Department of Physics and Astronomy, Uppsala University, Uppsala, Sweden}
\author{Keivan G. Stassun} 
\affil{Department of Physics and Astronomy, Vanderbilt University, Nashville, TN USA} 
\affil{Physics Department, Fisk University, Nashville, TN USA} 

\altaffiltext{1}{NSF REU Student at Vanderbilt University}

\begin{abstract}
We present the discovery of an extended ring of ultraviolet emission surrounding the AGB star U Hya in archival observations performed by the Galaxy Evolution Explorer (GALEX). 
This is the third discovery of extended UV emission from a carbon AGB star and the first from an AGB star with a detached shell.
From imaging and photometric analysis of the FUV and NUV images, we determined that the ultraviolet ring has a radius of $\sim 110^{\prime\prime}$, thus indicating that the emitting material is likely associated with the detached shell seen in the infrared.  
We find that scattering of the central point source of NUV and FUV emission by the dust shell is negligible. 
Moreover, we find that scattering of the interstellar radiation field by the dust shell can contribute at most $\sim10\%$ of the FUV flux. 
Morphological and photometric evidence suggests that shocks caused by the star's motion through space and, possibly, shock-excited H$_2$ molecules are the most likely origins of the UV flux.
In contrast to previous examples of extended UV emission from AGB stars, the extended UV emission from U Hya does not show a bow shock-like structure, which is consistent with a lower space velocity and lower interstellar medium density. 
This suggests the detached dust shell is the source of the UV emitting material and can be used to better understand the formation of detached shells.
\end{abstract}


\section{Introduction}

Detached dust shells are found around some asymptotic giant branch (AGB) stars (typically the carbon-rich type) and their origins are still a matter of debate. 
Some argue that detached shells indicate episodic mass loss events that lead to two-wind interactions \citep[e.g.,][]{2005A&A...436..633S}, others suggest detached shells are density enhancements following a temporary increase in the mass loss rate after a thermal pulse \citep[e.g.,][]{2007A&A...470..339M}. 
Additionally, the role of external forces, such as the interaction of the AGB wind with the interstellar medium (ISM), cannot be discounted \citep[e.g.,][]{2002ApJ...571..880V}. 
Theoretical studies of the formation of detached dust shells and their evolution in the ISM \citep{2000A&A...357..180S,2002ApJ...571..880V,2005A&A...436..633S,2006MNRAS.372L..63W,2007A&A...470..339M} target information provided by long wavelength observations. 
However, shocks and interactions with the ISM are better studied with short wavelength observations. 
Indeed, the Galaxy Evolution Explorer (GALEX) has revealed a bow shock-like structure ahead of the AGB star Mira plus a long trailing tail caused by the star's high space velocity through the ISM \citep{2007Natur.448..780M}. 
Targeted observations by GALEX of the carbon AGB star IRC+10216 uncovered the interaction of the AGB wind with the ISM in the star's astrosphere \citep{2010ApJ...711L..53S}. 
The most recent discovery of extended emission from AGB stars was reported by \citet{2014arXiv1408.1050S}, where deep GALEX observations imaged two half-ringed structures around the carbon AGB star CIT~6 in the FUV.  
However, with such few detections, the frequency and physical characteristics of extended UV emission is uncertain.

In this Letter, we present our discovery and analysis of  extended FUV and NUV emission from the carbon AGB star U Hya from GALEX All-Sky Imaging Survey (AIS) observations.
We find that extended FUV and NUV emission is spatially coincident with the detached dust shell initially discovered in the infrared with IRAS \citep{1994A&A...281L...1W} and recently studied with the AKARI satellite \citep{2011A&A...528A..29I} and the Herschel Space Observatory \citep{2012A&A...537A..35C}. 
We discuss the most likely origin of the UV emission and the questions posed by this discovery. 
Our discovery of  an extended ring of UV emission around U Hya marks the third discovery of extended UV emission from a carbon AGB star and the first of an AGB star with a detached shell.
This particular discovery of extended emission from shallow GALEX AIS exposures is remarkable given that previous examples required exposures that were two orders of magnitude longer, leading some authors to suggest that this type of extended emission would likely not be observable in short GALEX exposures \citep{2014arXiv1408.1050S}. 
Rather, we suggest that the existing archive of observations is a potential trove of similar detections in a much larger sample of AGB stars.

\begin{figure*}
\centering
\includegraphics[scale=0.55]{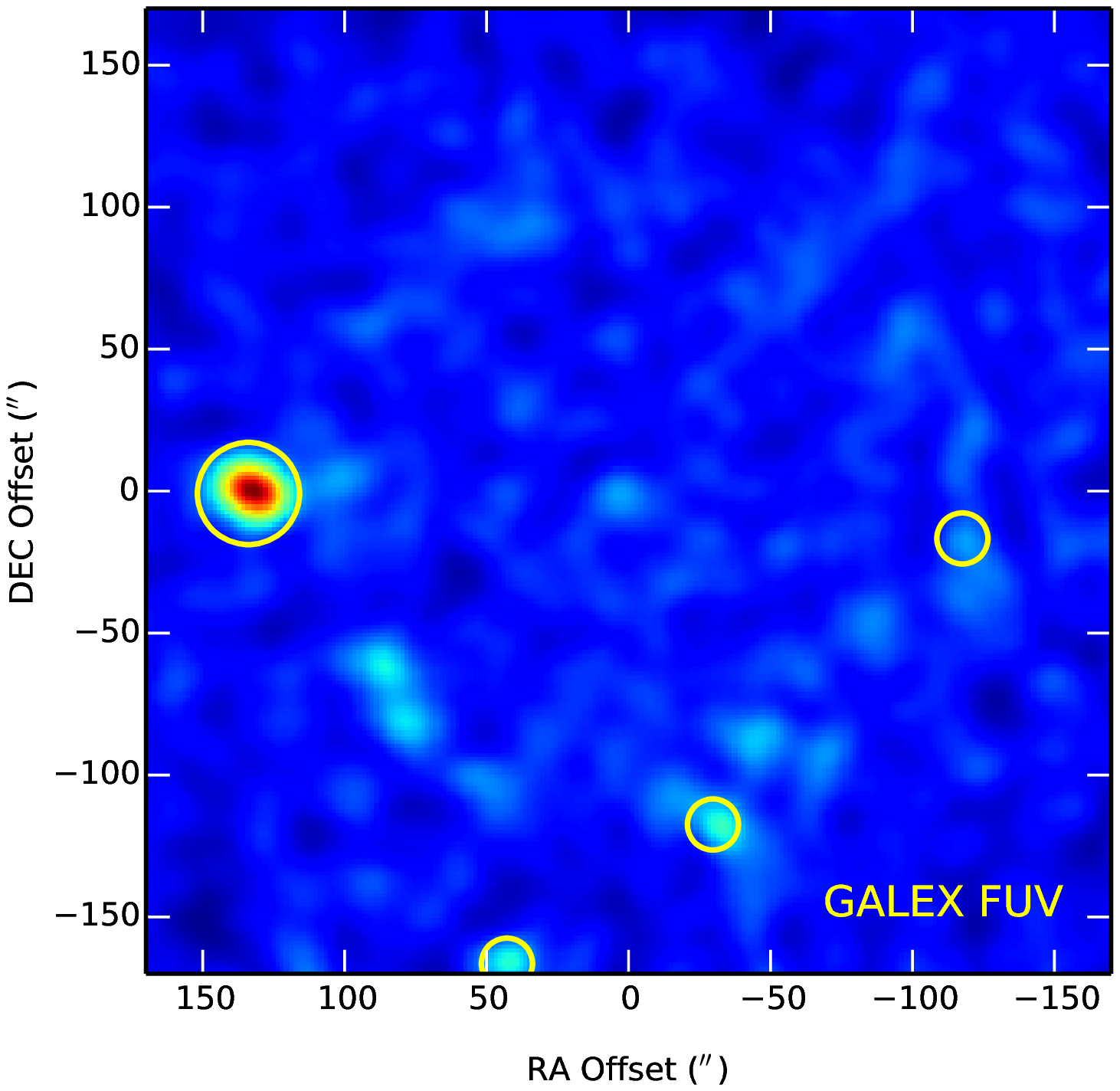} 
\includegraphics[scale=0.55]{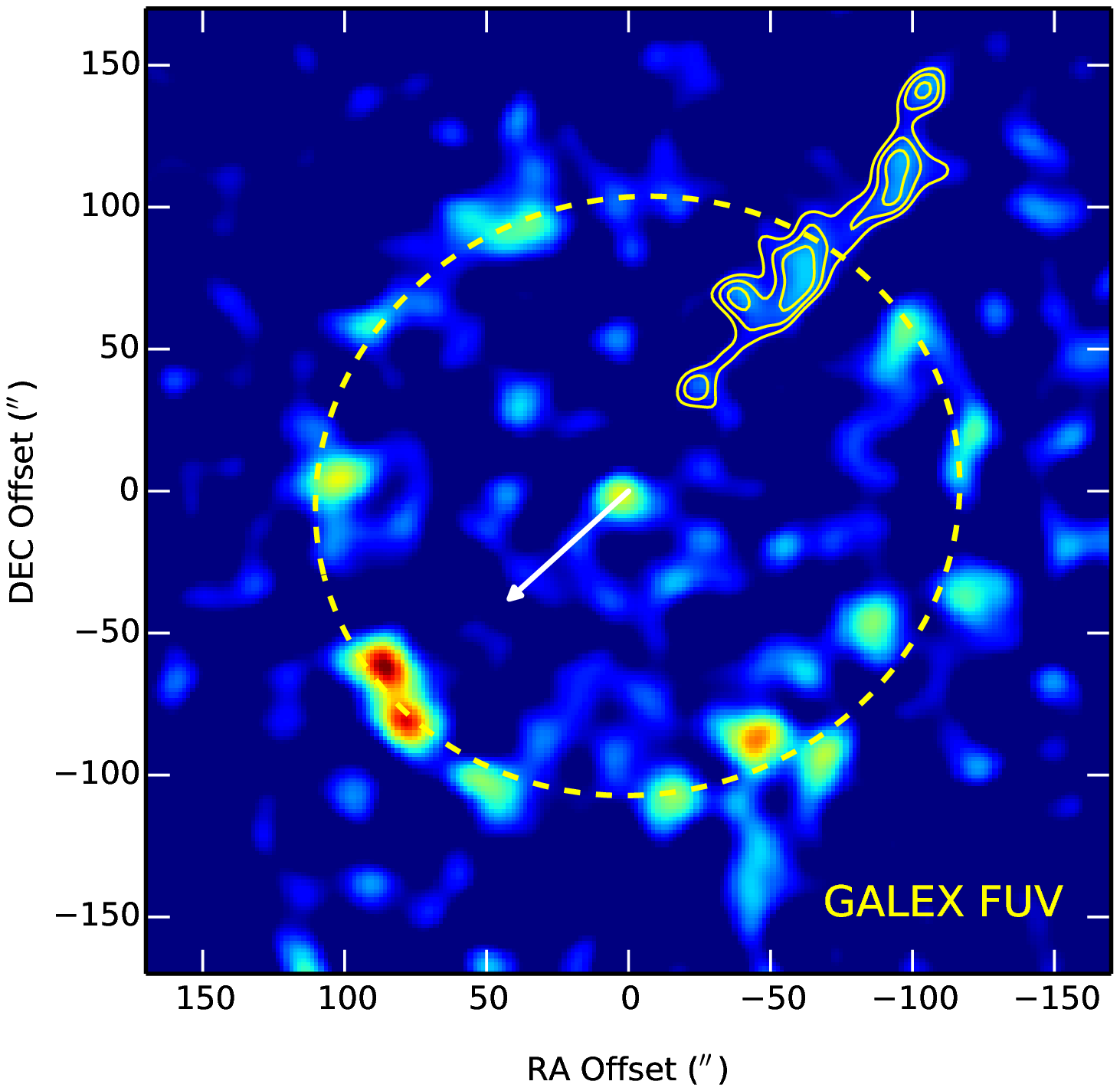} \\
\includegraphics[scale=0.55]{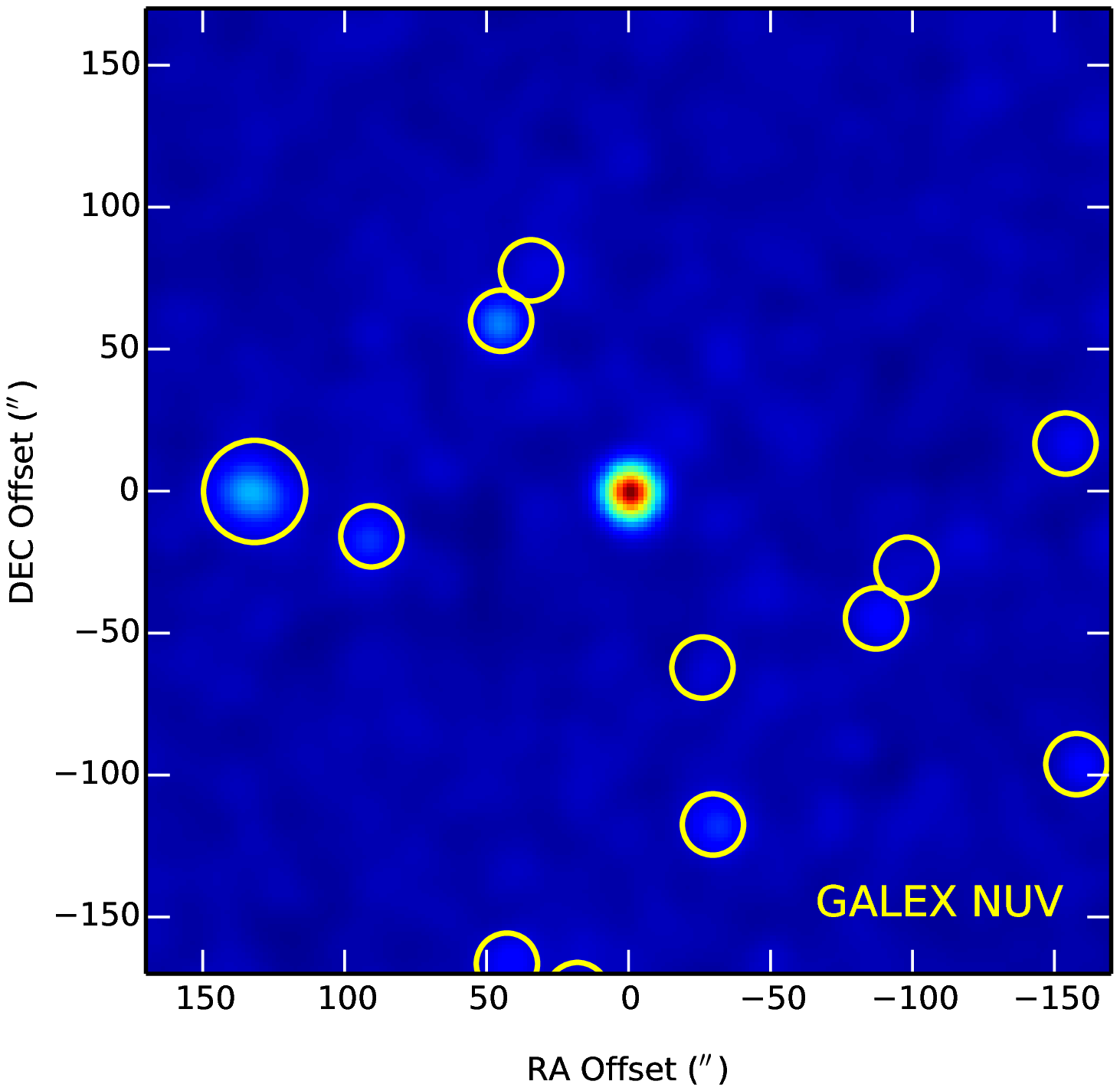}
\includegraphics[scale=0.55]{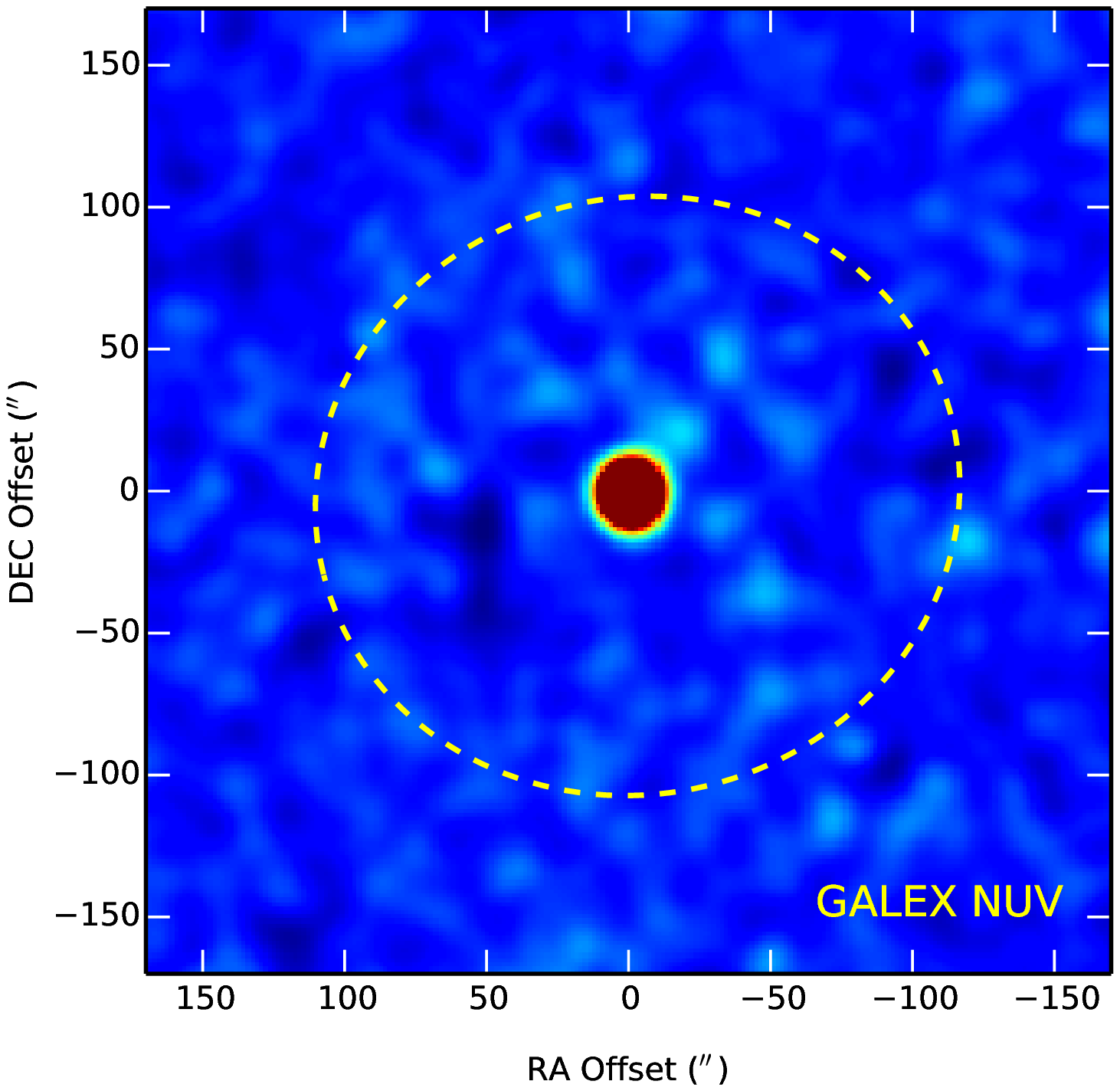} 
\caption{FUV (top panels) and NUV (bottom panels) images of U Hya as observed in the GALEX AIS images. The left panels show field source regions selected for removal. In the right panels, field sources have been removed and the best-fit ellipse to the bright FUV emission is depicted by the dashed line (see text). In the upper right panel, the faint contours of the spoke-like feature are included along with the proper motion vector. \label{fig1}}
\end{figure*}

\section{Observations}

Observations were performed by the GALEX satellite observatory \citep[see][]{2005ApJ...619L...7M} as part of the all-sky survey for 198 seconds in the NUV (1771-2831 \AA) and FUV (1344-1786 \AA) bandpasses on 26 March 2006\footnote{A deeper NUV Guest Investigator observation was performed but is compromised by a large artifact across U Hya.}. 
We retrieved the FUV and NUV images sampled at a resolution of  $1\farcs5$ per pixel from the GALEX archive using {\it GalexView}\footnote{http://galex.stsci.edu/GalexView/}.
The point spread function of the GALEX imaging system has a core FWHM of $\sim 5^{\prime\prime}$.  
The AGB star itself is apparently detected in both the FUV and NUV bands with {\it GalexView} catalog fluxes (provided in Table~\ref{fluxes}). 
In the FUV image, a faint ring of emission can be seen at $\sim 110^{\prime\prime}$, but this ring is not immediately apparent in the NUV image (Figure~\ref{fig1}). 

\section{Analysis\label{analysis}}

The FUV and NUV images required additional analysis to establish the existence and characteristics of the extended UV emission. 
As described in the following, we identified and removed likely foreground and background field sources, characterized the distribution of the UV emission by studying the bright emission spots, and measured the surface brightness radial profiles with respect to the AGB star. 

\subsection{Field Source Removal} 

We selected field sources for removal by identifying stellar and extragalactic counterparts to the UV point-like sources. 
We used images and source catalogs from the Two-Micron All Sky Survey (2MASS) to make the identifications. 
Pixels corresponding to the UV emission from stellar or extragalactic sources were replaced with a Poisson-distributed noise pattern drawn from a source-free region in the vicinity of the removed source. 
Images with and without field sources are presented in Figure~\ref{fig1}, where we have also smoothed the images with a 2D Gaussian kernel that has a FWMH of $\sim14^{\prime\prime}$. 
The extended FUV emission appears as a distribution of bright spots on a ring-like structure. 

\subsection{Distribution of FUV Emission \label{section_fuvdistribution}} 

We determined the distribution of the bright spots of FUV emission seen in Figure~\ref{fig1} by visually identifying the brightest spots of emission from $70^{\prime\prime}$ to $130^{\prime\prime}$.  
We found no compelling evidence for corresponding bright spots in the NUV image. 
Using the coordinates of these bright FUV spots we performed a least-squares fit to an ellipse with the semi-major/semi-minor axis, position angle, and the coordinates of the ellipse center as free parameters. 
A bootstrapping method is used to estimate the best-fit errors by adding normally-distributed noise ($\sigma \sim 5^{\prime\prime}$) to the bright spot coordinates and selecting 9 spots at random for each of the 10,000 least square fits. 
From the parameter statistics of the 10,000 trials we estimate that the best-fit ellipse has semi-major and semi-minor axes of $114\pm 8^{\prime\prime}$ and $105 \pm 10^{\prime\prime}$, respectively, and the semi-major axis has a position angle of $103^{\circ} \pm 24^{\circ}$. 
The uncertain position angle of the semi-major axis is approximately consistent with the proper motion vector of U Hya  \citep[$\mu_a = 42.6 {\rm ~mas~yr}^{-1}$ and $\mu_d = -37.7 {\rm ~mas~yr}^{-1}$,][]{2007A&A...474..653V}, suggesting the ellipsoidal shape may be due to the shell's motion, while the misalignment may suggest asymmetries in the ISM. 
The best-fit ellipse is overlaid upon the FUV and NUV images along with the proper motion vector of U Hya in Figure~\ref{fig1}). 

The best-fit ellipse centroid is offset from the celestial coordinates of U Hya by $3\farcs7$ ($\delta[{\rm RA},{\rm Dec}] = [-3\farcs1, -2\farcs0]$), which is within the ellipse centroid error estimate of $\sim 4^{\prime\prime}$.
However, the central FUV point source location is less consistent with the celestial position of U Hya than the brighter central NUV point source. 
The FUV central point source is offset from the celestial coordinates of U Hya by $5\farcs1$ ($\delta[{\rm RA},{\rm Dec}] = [+5\farcs0, -1\farcs0]$) and $5\farcs4$ from the ellipse centroid.  
Given the GALEX astrometric accuracy of $\sim0\farcs5$, the offset of the FUV central source is significant and considered further in \S\ref{section_peculiar}.

As a final remark on the distribution of the FUV emission, we note a faint incomplete radial ``spoke"-like feature at a position angle between $310^{\circ}$ to $330^{\circ}$ (Figure~\ref{fig1}). 
The integrated background-subtracted flux of the spoke-like feature is $(3.6\pm1.4)\times10^{-5}~\photonfluxunit$, hence, it is only tentatively detected.
However, the position angle of this feature is opposite the star's proper motion suggesting the potential association with the past position of U Hya (see \S\ref{section_peculiar}). 

\begin{figure}[ht]
\centering
\includegraphics[scale=0.55]{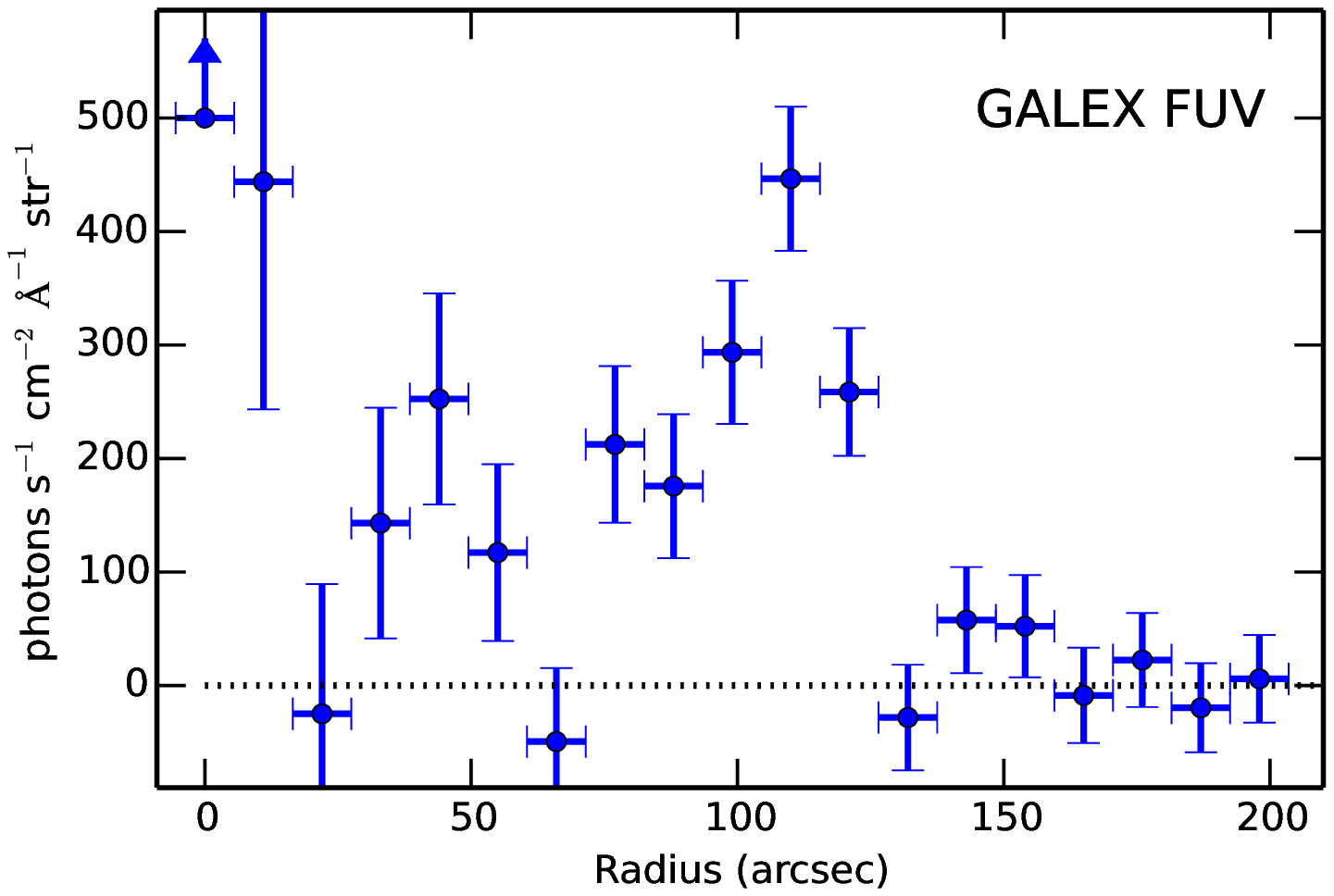} 
\includegraphics[scale=0.55]{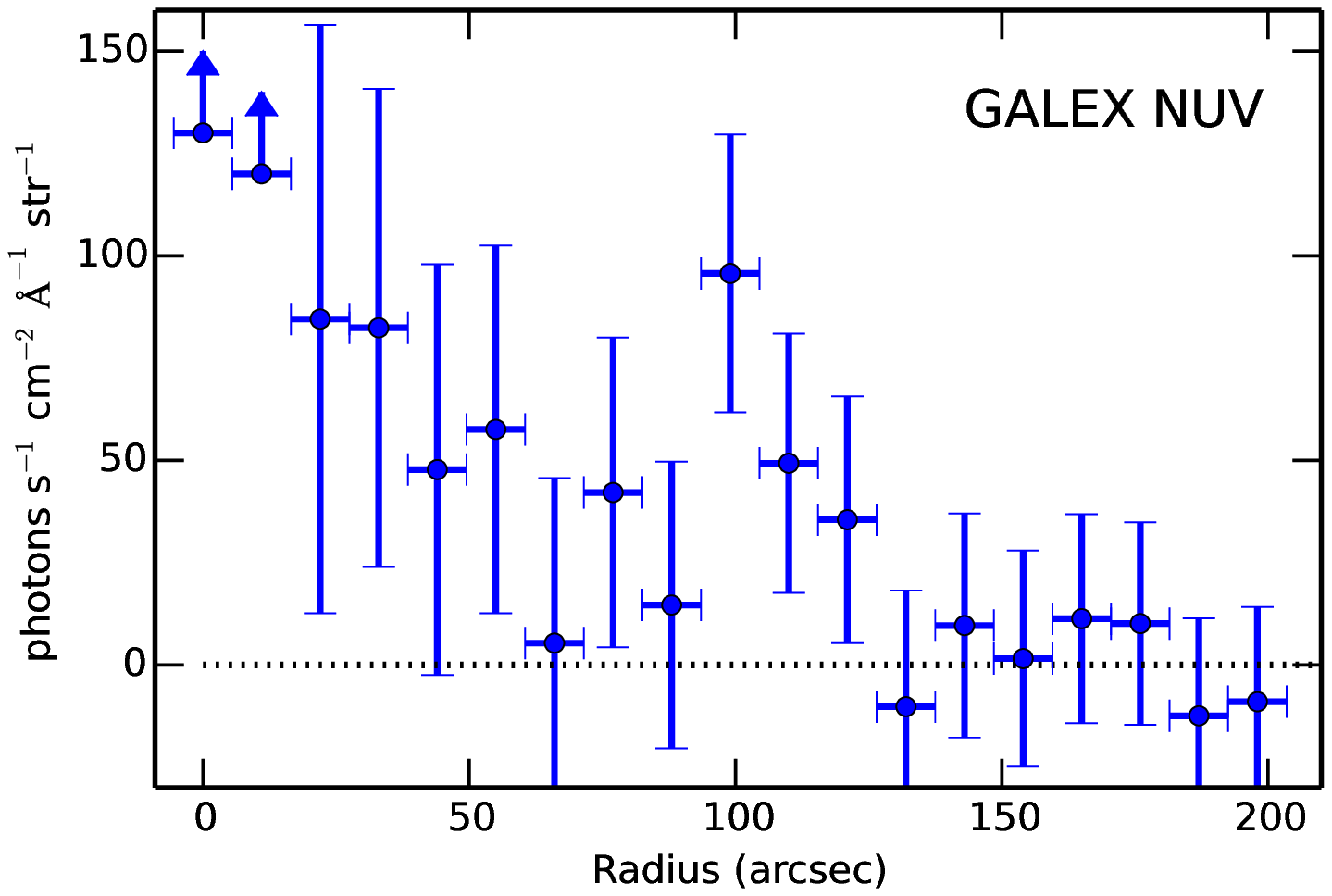} 
\caption{FUV (top panel) and NUV (bottom panel) background-subtracted surface brightness radial profiles. 
Surface brightness axis is limited to highlight the extended emission near $\sim100^{\prime\prime}$, bins closest to the central point source emission are indicated with upward pointing arrows because their emission lies beyond the plot range.\label{fig2}}
\end{figure}

\subsection{Surface Brightness Radial Profiles \label{section_profiles}}

We extracted surface brightness measurements from 18 nested circular annuli and one central circular aperture for both FUV and NUV images.
Each annulus is centered on the celestial coordinates of U Hya and has a width of $11^{\prime\prime}$, the central circular aperture has a radius of $11^{\prime\prime}$. 
The four farthest annuli (from $160^{\prime\prime}$ to $203^{\prime\prime}$) were used to estimate the total background (sky and detector) surface brightness. 
We estimate the FUV and NUV background surface brightness is $4.25\times10^{-4} {\rm ~counts~s}^{-1} {\rm ~pixel}^{-1}$ and $3.10\times10^{-3} {\rm ~counts~s}^{-1} {\rm ~pixel}^{-1}$, respectively.
For each annulus we use the background-subtracted counts to estimate the source Poisson error then add this in quadrature to the background error to determine the source error. 
We converted the total background-subtracted count rates (CPS) to fluxes using the relationships from \citet{2005ApJ...619L...7M} and the GALEX online documentation.\footnote{\url{http://galexgi.gsfc.nasa.gov/docs/galex/instrument.html}} 

The resulting FUV and NUV background-subtracted surface brightness radial profiles are shown in Figure~\ref{fig2}. 
The radial profiles begin at the position of the stellar point source and reach a maximum between $100^{\prime\prime}$ and $130^{\prime\prime}$, with the peak at $\sim110^{\prime\prime}$.
At a distance of 208 pc, this peak corresponds to a radius of $\sim0.11$~pc from the celestial position of U Hya.
The radial profile behavior is consistent with a shell-like structure that forms a ring when projected onto the sky. 
Although the ring is not apparent in the NUV image, it is revealed in the surface brightness radial profile (Figure~\ref{fig2}). 

In the FUV radial profile, a single radial bin near $\sim 45^{\prime\prime}$ rises $\geq2\sigma$ above the background. 
Examination of this radius in Figure~\ref{fig1} suggests a faint semi-circular arc in the southern half of the FUV image. 
There is no clear evidence for a similar feature in the NUV image nor in the NUV radial profile. 
However, the broad shoulders of the NUV imaging PSF extend to $\sim50^{\prime\prime}$, making the identification of such a peak more difficult in the high background image.

We have calculated the total background-subtracted FUV and NUV fluxes in these rings by integrating the surface brightness flux from $27\farcs5$ to $60\farcs5$ for the inner ring and $71\farcs5$ to $126\farcs5$ for the outer ring and associated errors added in quadrature. 
Variations within the annuli, like the bright spots studied in \S\ref{section_fuvdistribution}, are not included in the error budget. These variations may be due to statistical fluctuations but could also indicate clumps of emitting material (see \S\ref{section_h2}). 
The total FUV and NUV fluxes for these shells are reported in Table~\ref{fluxes}.

\subsection{Location of UV emission with respect to the IR-emitting dust shell}

To identify the spatial relationship of the UV emission with respect to the IR-emitting  dust shell, we overlaid contours of the FUV emission onto the 70$\mu m$ {\it Herschel} Photoconductor Array Camera and Spectrometer (PACS) image \citep[see][]{2011A&A...526A.162G}.
The 70$\mu m$ PACS imaging PSF is comparable to that of GALEX ($\sim 5^{\prime\prime}$). 
The best-fit ellipse and location of the FUV emission (Figure~\ref{fig3}) suggest that the UV is coincident, within astrometric and resolution uncertainties, with $70\mu m$ dust emission (Figure~\ref{fig3}).

\begin{deluxetable}{lr}
\tablecolumns{2}
\tablewidth{220pt}
\tablecaption{GALEX UV Properties for U Hya \label{fluxes}}
\tablehead{
\colhead{Parameter} & \colhead{Value\tablenotemark{a}} }
\startdata
\sidehead{Point Source Flux Densities}

\hspace{0.5cm} $f_{\rm cs, FUV}$ & $\csfuvflux$ \\ 
\hspace{0.5cm} $f_{\rm cs, NUV}$ & $\csnuvflux$ \\
 
\sidehead{Dust Shell Flux Densities} 

\hspace{0.5cm} $f_{\rm outer, FUV}$ & $\orfuvflux$ \\
\hspace{0.5cm} $f_{\rm outer, NUV}$ & $\ornuvflux$ \\
\hspace{0.5cm} $f_{\rm inner, FUV}$ & $\irfuvflux$ \\  

\sidehead{Dust Shell Scattering Fractions\tablenotemark{b}} 

\hspace{0.5cm} Point Source & $\approx 1\%$ \\ 
\hspace{0.5cm} ISRF of FUV & up to 10\% \\ 
\hspace{0.5cm} ISRF of NUV & up to 100\% \\

\enddata
\tablenotetext{a}{Flux densities are provided in $\photonfluxunit$. Stellar point source fluxes acquired from GALEX archive while dust shell fluxes and uncertainties are discussed in \S\ref{section_profiles}.} 
\tablenotetext{b}{Based on infrared properties from \citet{2011A&A...528A..29I} and \citet{2012A&A...537A..35C} and the formulation in \citet{2001A&A...372..885G} as discussed in \S\ref{section_scattering}.}
\end{deluxetable}

\section{Origin of the Extended UV Emission}

We trace the origin of the extended UV emission by considering the most likely scenarios that could explain the UV emission.
The physical processes considered are scattering of UV photons, shocks formed when the AGB wind encounters the ISM, shocks from the star's motion through the ISM, or, excitation of molecular hydrogen.
These processes are not mutually exclusive and they may all occur simultaneously to some degree. 

\subsection{Scattering by the Detached Dust Shell\label{section_scattering}}

Given the spatial relationship of the IR and UV emission (Figure~\ref{fig3}), we consider the possibility that UV photons are scattered by the dust shell into our line of sight. 
Scattering of this type is seen in optical observations of numerous AGB stars with dust shells \citep{2001A&A...372..885G,2010A&A...515A..27O,2011A&A...531A.148R} including U Hya \citep{2007ASPC..378..305I}. 
The scattering efficiency increases at shorter wavelengths and the scattered flux and distribution can be used to infer the dust shell properties  \citep[e.g., shell mass, $M_{\rm shell}$, and shell radius, $R_{\rm shell}$, see][]{2001A&A...372..885G}. 
However, in our case, we used this formulation \citep{2001A&A...372..885G} and IR dust shell properties\citep{2011A&A...528A..29I} to estimate the fraction of central FUV and NUV flux scattered by the dust shell (see Table~\ref{fluxes}).
We conclude that scattering of the central source by the dust shell can only account for $1\%$ of the extended UV emission.

Scattering of the interstellar radiation field (ISRF) by the dust shell is another potential source of the extended UV emission from U Hya.
The expected contributions from FUV and NUV flux densities (in $\photonfluxunit$) due to scattering of the ISRF by dense regions of the ISM is roughly equal due to the respective bandpasses and the UV extinction curve \citep{2009ApJ...692.1333S}. 
If this proportionality holds in the dust shell of U Hya, then assuming all the NUV flux from the dust shell arises from scattering of the ISRF, we can only account for $\sim10\%$ of the FUV flux.
Hence, it is unlikely that the FUV flux is caused by the scattering processes alone. 
Instead, we suggest that shocks are responsible for some or all of the FUV and NUV emission from the detached dust shell of U Hya. 

\subsection{Shocks from the Shell, ISM, and Proper Motion\label{section_shocks}}

Our next clue that shocks might be responsible for the extended FUV and NUV emission can be seen in Figures~\ref{fig1}\&\ref{fig3}, where the FUV ring is brightest in the direction of the star's proper motion. 
The peculiar velocity of U Hya \citep[$V_{*}\sim72 {\rm ~km~s}^{-1}$;][]{2012A&A...537A..35C}, is capable of producing post-shock temperatures of $\sim1.6\times10^{5} {\rm ~K} (\mu_{H}/1.33) (V_{*}/72 {\rm ~km~s}^{-1})$, for strong shock conditions and a typical mean molecular mass, $\mu_{H}$.
Such a high-temperature shock is consistent with the increasing flux towards shorter wavelengths. 
Similar characteristics are seen in GALEX image of IRC+10216, where the relative motion of the star with respect to the ISM forms a strong shock with a post-shock temperature of about $T_{\rm s} \sim 10^{5} {\rm ~K}$ \citep{2010ApJ...711L..53S}. 
However, unlike IRC+10216, where the extended NUV emission leads the FUV emission,  in U Hya the FUV and NUV extended emission are coincident with tentative evidence that the FUV emission actually leads the NUV emission. 
But the larger distance to U Hya and short exposure time (more than two orders of magnitude shorter than the observation of IRC+10216) make it difficult to determine the precise spatial relationship of the UV emission. 
We note, however, that there is no apparent bow shock structure expected from the interaction with the ISM in the PACS images \citep{2012A&A...537A..35C}. 
This is likely due to a combination of the slower peculiar velocity of U Hya and a lower ISM density. 

\subsection{Molecules Surviving in Clumps? \label{section_h2}} 

Alternatively, collisional excitation of H$_2$ by hot electrons in a low-density shocked gas might explain the enhancement of the FUV emission, similar to the long-lived FUV tail of Mira \citep{2007Natur.448..780M}. 
The molecular composition of the detached shell around U Hya is poorly-studied. 
\citet{1994A&A...281L...1W} argued that the dynamical age of the detached dust shell suggests that despite self-shielding, CO molecules should be disassociated by now due to the ISRF.
The survival of H$_2$ against disassociation by the ISRF is less clear but the typically higher abundance of H$_2$ suggests it could survive after all the CO has been disassociated. 
If both CO and H$_2$ molecules have survived disassociation by the ISRF, they are likely to do so in clumps like those seen in scattered light around the detached shell of R Scl and U Cam \citep{2010A&A...515A..27O}. 
Such clumps would be consistent with the observed morphology of the FUV emission as a series of bright spots on an elliptical ring, but the low-signal in our shallow exposures limit our ability to assert a clumpy environment.

\subsection{Peculiar features of the FUV emission \label{section_peculiar}}

\begin{figure}
\centering
\includegraphics[scale= 0.55]{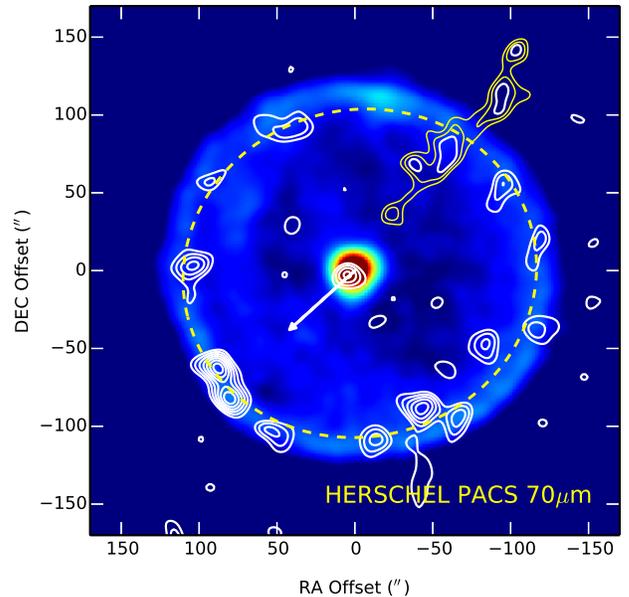}
\caption{Dust shell of U Hya observed by Herschel PACS in 70$\mu$m band with contours of the GALEX FUV emission overlaid. 
The yellow contours correspond to the faint spoke-like feature discussed in \S\ref{section_peculiar}.
The proper motion vector is indicated along with the best-fit ellipse for the bright FUV points. \label{fig3}}
\end{figure}

In \S\ref{section_fuvdistribution}, we highlighted two peculiar features of the FUV emission, the apparent offset of the FUV central point source from the celestial position of U Hya by $\sim5^{\prime\prime}$ and the faint spoke-like feature opposite the star's proper motion. 
Since the bright NUV central point source coincides with the celestial coordinates of U Hya within GALEX astrometric uncertainty, we tentatively conclude that the offset of the central FUV source from the star is real. 
This offset might suggest a heretofore unknown distant (up to 1000 AU) hot companion, similar to that suggested for other GALEX detections of AGB stars \citep{2008ApJ...689.1274S}. 
However, because the offset lies in the direction of U Hya's motion (see Figure~\ref{fig3}), it may suggest shocks caused by the star's motion. 
But the presence of a detached shell implies there is no ISM near the star, so such shocks would have to arise from an exotic scenario where the star plows into its own wind. 
In the opposite direction of the proper motion, the spoke-like feature seen in the FUV is reminiscent of Mira's long trailing tail \citep{2007Natur.448..780M}, albeit a smaller version. 
Mira's tail covers 30,000 years of mass loss where as the purported tail of U Hya covers only $\sim2,000$ years assuming its length is $\sim150^{\prime\prime}$. 
If confirmed, this tail of emission can be used to constrain the mass loss history since the formation of the dust shell that has a dynamical age of $\sim$8,000 years. 
We recommend further study to confirm the origin and extent of these features from U Hya.

\section{Conclusions}

Analysis of archival GALEX AIS observations of U Hya establishes the detection of extended ultraviolet emission in the NUV and FUV for the first time in an AGB star with a detached dust shell and only the third of this type of emission from a carbon AGB star. 
The FUV the emission appears as a series of bright spots that together form a ring-like structure similar in size to that of the IR-emitting detached dust shell (Figures~\ref{fig3}).
The radius is approximately $110^{\prime\prime}$, but is better described by an ellipse with a semi major axis of $114^{\prime\prime}$ and semi minor axis of $105^{\prime\prime}$, the semi-major axis appears projected onto the plane of the sky with a position angle of $\sim100^{\circ}$.
In the NUV, the extended emission is not apparent in the image (Figure~\ref{fig1}) but is recovered in surface brightness profiles (Figure~\ref{fig2}). 
Based on these profiles, the extended NUV emission has a similar, if not smaller, with a radius of approximately $106^{\prime\prime}$. 

We measured the total flux from the surface brightness profiles and compared it to that expected from scattering by the dust shell using the dust shell properties derived from  infrared observations.
We conclude that scattering by the central star is insufficient to explain the UV emission and scattering by the ISRF cannot be ruled out.
However, assuming all the NUV arises from scattering of the ISRF, $\sim90\%$ of the FUV flux remains unexplained.
Based on the peculiar velocity of U Hya and the enhanced flux in the direction of the star's proper motion, we conclude that some or all of the NUV flux and most of FUV flux arises from shocks caused by the star's motion through the ISM. 

This discovery of extended UV emission from U Hya, along with that of carbon stars IRC+10216 and CIT 6, leads one to speculate if this type of UV emission is a common trait of the circumstellar envelopes around carbon AGB stars, high-velocity AGB stars, or inherent to the formation of the detached shell. 
Other than their classification as carbon AGB stars, these three sources of extended UV emission are dissimilar. 
A thorough survey of GALEX UV observations of AGB stars that utilizes the techniques developed here is underway to establish the origin(s) and frequency of this type of UV emission from these cold, dusty environments.

\acknowledgments

E. S. wishes to acknowledge the NSF REU site grant PHY-1263045 to Vanderbilt University and the National Society of Hispanic Physicists' Victor M. Blanco Fellowship.
R. M. wishes to acknowledge the Henri Chr\'etien International Research Grant awarded by the American Astronomical Society that facilitated collaboration on this research program. 

{\it Facilities:} \facility{GALEX, Herschel}.

\vfill \hfill End of accepted material.

\begin{figure*}
\centering
\includegraphics[scale= 0.85]{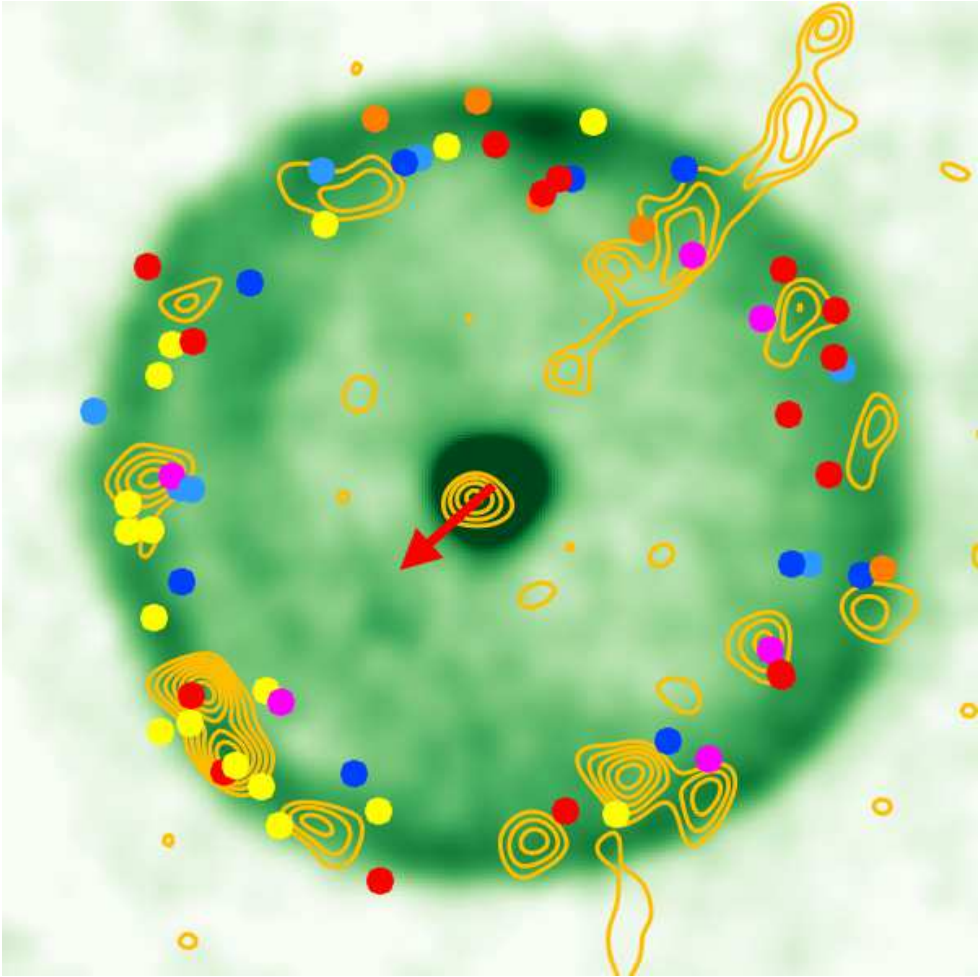}
\caption{{\it Holiday Edition of Figure 3}. Visit http://www.rudyphd.com/uhyawreath.gif for animated version.}
\end{figure*}


\begin{thebibliography}{}
\bibitem[Cox et al.(2012)]{2012A&A...537A..35C} Cox, N.~L.~J., Kerschbaum, F., van Marle, A.-J., et al.\ 2012, \aap, 537, A35 
\bibitem[Gonz{\'a}lez Delgado et al.(2001)]{2001A&A...372..885G} Gonz{\'a}lez Delgado, D., Olofsson, H., Schwarz, H.~E., Eriksson, K., \& Gustafsson, B.\ 2001, \aap, 372, 885 
\bibitem[Groenewegen et al.(2011)]{2011A&A...526A.162G} Groenewegen, M.~A.~T., Waelkens, C., Barlow, M.~J., et al.\ 2011, \aap, 526, A162 
\bibitem[Izumiura et al.(2007)]{2007ASPC..378..305I} Izumiura, H., Nakada, Y., Hashimoto, O., Mito, H., \& Hayashi, T.\ 2007, Why Galaxies Care About AGB Stars: Their Importance as Actors and Probes, 378, 305 
\bibitem[Izumiura et al.(2011)]{2011A&A...528A..29I} Izumiura, H., Ueta, T., Yamamura, I., et al.\ 2011, \aap, 528, A29 
\bibitem[Martin et al.(2007)]{2007Natur.448..780M} Martin, D.~C., Seibert, M., Neill, J.~D., et al.\ 2007, \nat, 448, 780 
\bibitem[Mattsson et al.(2007)]{2007A&A...470..339M} Mattsson, L., H{\"o}fner, S., \& Herwig, F.\ 2007, \aap, 470, 339 
\bibitem[Morrissey et al.(2005)]{2005ApJ...619L...7M} Morrissey, P., Schiminovich, D., Barlow, T.~A., et al.\ 2005, \apjl, 619, L7 
\bibitem[Olofsson et al.(2010)]{2010A&A...515A..27O} Olofsson, H., Maercker, M., Eriksson, K., Gustafsson, B., \& Sch{\"o}ier, F.\ 2010, \aap, 515, A27 
\bibitem[Ramstedt et al.(2011)]{2011A&A...531A.148R} Ramstedt, S., Maercker, M., Olofsson, G., Olofsson, H., \& Sch{\"o}ier, F.~L.\ 2011, \aap, 531, A148
\bibitem[Sahai \& Mack-Crane(2014)]{2014arXiv1408.1050S} Sahai, R., \& Mack-Crane, G.~P.\ 2014, arXiv:1408.1050 \bibitem[Sahai \& Chronopoulos(2010)]{2010ApJ...711L..53S} Sahai, R., \& Chronopoulos, C.~K.\ 2010, \apjl, 711, L53 
\bibitem[Sahai et al.(2008)]{2008ApJ...689.1274S} Sahai, R., Findeisen, K., Gil de Paz, A., \& S{\'a}nchez Contreras, C.\ 2008, \apj, 689, 1274 
\bibitem[Sch{\"o}ier et al.(2005)]{2005A&A...436..633S} Sch{\"o}ier, F.~L., Lindqvist, M., \& Olofsson, H.\ 2005, \aap, 436, 633 
\bibitem[Steffen \& Sch{\"o}nberner(2000)]{2000A&A...357..180S} Steffen, M., \& Sch{\"o}nberner, D.\ 2000, \aap, 357, 180 
\bibitem[Sujatha et al.(2009)]{2009ApJ...692.1333S} Sujatha, N.~V., Murthy, 
J., Karnataki, A., Henry, R.~C., \& Bianchi, L.\ 2009, \apj, 692, 1333 
\bibitem[van Leeuwen(2007)]{2007A&A...474..653V} van Leeuwen, F.\ 2007, \aap, 474, 653 
\bibitem[Villaver et al.(2002)]{2002ApJ...571..880V} Villaver, E., Garc{\'{\i}}a-Segura, G., \& Manchado, A.\ 2002, \apj, 571, 880 
\bibitem[Wareing et al.(2006)]{2006MNRAS.372L..63W} Wareing, C.~J., Zijlstra, A.~A., Speck, A.~K., et al.\ 2006, \mnras, 372, L63 
\bibitem[Waters et al.(1994)]{1994A&A...281L...1W} Waters, L.~B.~F.~M., Loup, C., Kester, D.~J.~M., Bontekoe, T.~R., \& de Jong, T.\ 1994, \aap, 281, L1 
\end{thebibliography}
\end{document}